### Forecasts of Cancer & Chronic Patients
*Big Data Metrics of Population Health*

Jacob Kuriyan and Nathaniel Cobb

## Abstract

Chronic diseases and cancer account for over 75% of health care costs in the US. Increased prevention services and improved primary care are thought to decrease costs. Current models for detecting changes in the health of populations are cumbersome and expensive, and are not sensitive in the short term. In this paper we model population health as a dynamical system to predict the time evolution of the new diagnosis of chronic disease and cancer. This provides a reliable forecasting tool and a means of measuring short-term changes in the health status of the population resulting from preventive care programs. 12 month forecasts of cancer and chronic populations were accurate with errors lying between 3% and 6%. We confirmed what other studies have demonstrated that diabetes patients are at increased cancer risk but, interestingly, we also discovered that all of the studied chronic conditions increased cancer risk just as diabetes did, and by a similar amount. The model (i) yields a new metric for measuring performance of preventive and clinical care programs that can provide timely feedback for quality improvement programs; (ii) helps understand "savings" in the context of preventive care programs and explains how they can be calculated in the short term even though they materialize only in the long term and (iii) provides an analytic tool and metrics to infer correlations and derive insights on the effectiveness of care programs in improving health of populations and lowering costs.

## Introduction

Healthcare costs in the US are huge and at over 18% of GDP they are threatening the nation's economy. Over 75% of these costs can be attributed to treatment of chronic and cancer patients and so it is clear that any serious cost reductions will require a better understanding of the root causes of these diseases and implementing effective preventive care programs. Currently smoking and obesity are recognized as risk factors underlying chronic diseases and cancer and scientists are busy identifying other correlations to explain the biological mechanisms and relationships of these diseases.

Most of these studies (1, 2, 3,4,5,6,7) are epidemiological and they are laborious and made more difficult because they require medical data of populations that at this time are mostly available on paper records. In addition, their focus is usually on single diseases, and they leave open the issue of how ignoring co-morbid conditions (8) may affect their conclusions.

In the early days, economists plumbed vast amounts of data to search for correlations so as to understand how the market works. But very soon they started to assemble their "big data" consisting of census, labor and other economic data into the repositories of the bureau of labor statistics and devised sophisticated models to explain the correlations and also forecast changes in market conditions. US healthcare is also starting to collect data from disparate sources so as to harness the power of big data. Life sciences have seen the benefits of big data in their genome research and big data will play a crucial role in the recently announced brain mapping program. Pharma companies have similarly benefited from the genetic research results contained in big data in their effort to develop more efficient and targeted drugs. Archimedes, Inc., is a pioneering healthcare company that offers modeling and analytic solutions in healthcare, where mashed-up data from electronic health records, claims, clinical trials and pharmaco-chemistry can be used with their proprietary algorithms to simulate outcomes.

Big data for population health includes, besides the health of populations, social determinants such as the circumstances in which people are born, grow up, live, work, and age, as well as the systems put in place to deal with illness. These circumstances are in turn shaped by a wider set of forces: economics, social policies, and politics. Our focus will be in the subset of the big data that deals with the assessment of health of populations. Currently we have billions of healthcare (both medical and pharma) claims data stored in a national standard electronic format in the data warehouses of large insurers. Besides accurate demographic data, diagnosis and procedure codes are uniformly coded in the claims, and the diagnoses represent the physician's summary determination of the health status of the patient. Prescription drug information is also available in the pharma claims, and may be used to validate the diagnoses. Insurance claims directly links diagnoses with costs. Each patient generates multiple independent claims and it is possible to perform integrity checks, so as to isolate and even correct miscoded data. One immediate challenge is to harness the power of big data so as to manage health of populations and reduce variations in care.

We accomplish this by developing a new and patent pending model (9) of a population to provide insight into the competitive effectiveness of preventive care programs and the time evolution of chronic & cancer populations. The central proposition underlying the model is a simple one. As healthy people age, they acquire chronic diseases and for reasons unspecified in the model some go on to develop cancer. Our model attempts to simulate these changes that take place in the health of populations over time. One of our goals is to forecast the distribution of chronic and cancer patients in a population as well as their impact on healthcare costs.

Population distribution forecasts are attractive for financial and clinical reasons. Rising healthcare costs are a major concern to the nation and since 75% or more of these costs are attributed to chronic and cancer patients (10), being able to forecast their numbers accurately will help communities better budget their resources. Clinicians needed to support the aging population are also in short supply and forecasts can help distribute resources optimally. Further, prevention programs are expensive and their effectiveness is only evident a decade or more later, making it too late to correct mistakes. In contrast, forecasts based on pre-intervention data can be matched against post-intervention results to verify that preventive care programs are working as intended and if not, timely corrections can be introduced.

We will model a population as a dynamical system. There are numerous examples of dynamical system models in physics and engineering. Newton's explanation of planetary motions and orbits, for example, is an early and successful application of the method of dynamical systems. The classic predator-prey models in ecology and biology are examples of a population modeled as a dynamical system (11). Another example, and one more similar to our approach is the SIR model in epidemiology that attempts to explain the spread of infectious diseases over time (12, 13, 14, 15). It predicts the changes that take place in the susceptible (S), infected (I) and recovered (R) populations during an epidemic and after preventive care is introduced.

Distinct from dynamical system models are the popular actuarial models that insurers use in healthcare. These have a statistical basis and use regression analysis to correlate dependent and independent variables at a selected point in time. Time dependence is absent in this formulation, so there is no mathematical basis to use them for forecasts. Still, they are used for predictions, based on the hope that the future will be a repetition of the past and the expectation that correlations between variables will continue as before. When systems do not change or change slowly, actuarial models work very well.

Contrasted against regression models, the time evolution of the variables in a dynamical system are explicitly described using a mathematical equation. The term "deterministic" is also used to describe the model because given the initial state, it will determine or forecast every future state of the system.

We start with a very simple version of a population as a dynamical system to illustrate the method and establish the general validity of the approach. Next we consider a more elaborate model and use it to predict population distributions and the costs that will be incurred by both chronic and cancer patients. To connect with the more conventional analyses of epidemiologists, we also show how the model can analyze a population with a focus on a single disease and duplicate one of the published results on diabetes and cancer risk.

In our case the model serves two purposes. First, when solved, the equations will yield forecasts of chronic and cancer patients. Second, and even without obtaining a solution, the model offers a phenomenological language to describe the manner in which preventive care affects chronic populations, both in terms of costs and outcomes. At this time there are no easy ways of evaluating the performance of preventive care and clinical practice improvement programs, short of waiting for a decade to see the results. Our model yields a new metric for inferring that the care programs are effective in the short term. An observed deviation from the predicted state is then attributed to some new influence, such as an effective prevention program or a change in risk behavior. Phenomenology underlying the model also leads to a better understanding of why the premise of "shared savings" programs, promoted heavily by insurers and the federal government, may be flawed. It also becomes clear why we need to adopt a different paradigm for rewarding providers and cost justify investments in preventive care programs.

Note that throughout this paper, we use the terms "preventive care" and "prevention programs" to include both primary prevention (such as tobacco control and dietary interventions) and higher orders of prevention, such as good control of hypertension to prevent cardiovascular complications.

The paper is arranged as follows:  In section 2 we consider the three healthstate population model and derive the difference equation connecting a state vector at time "t" to time "t+1" through a transition matrix $M$.  In section 3 we describe the characteristics of the matrix $M$ and the physical meaning of the matrix elements in terms of the performance of preventive care programs.  In section 4 we solve the difference equation under the assumption that the matrix $M$ is time independent and forecast how the non-chronic, chronic and cancer populations evolve in time and compare them to observations obtained from claims data.   In section 5 we consider the more complex multiple healthstate model, with chronic patients segmented into 4 different healthstates of increasing comorbidities and severity of disease.  Once again we see that the forecasts of cancer and chronic patients match observations.   In section 6 we use the model to study single diseases and we investigate the impact of many different single chronic conditions on cancer risk.   In section 7 we show that "savings" in the context of preventive care programs must be interpreted with care and why the "shared savings" programs may not be a practical option for providers.  The concluding discussions are in section 8.

## *Methods and Results*

### 1. **The three healthstate population model**

We first consider a model where populations are segmented into three classes, ordered according to the severity of disease: N is the number of "non chronic" patients, meaning those without any of the seven chronic conditions listed below and they occupy the least severe healthstate; D is the number of patients with one or more of the listed chronic diseases and they occupy a more severe healthstate; and C is the number of cancer patients which we consider to occupy the most severe healthstate. As time passes, some people develop chronic conditions and others develop cancer, and they will move from one healthstate to another. The objective of the model is to calculate and forecast the count of populations in the three classes after a set period of time.

Typically in a sample population there could be hundreds of chronic conditions represented but to keep the analysis manageable we needed to restrict the number of them included in D. We used cost of treatment available in claims data to select the conditions to be included in D. We considered only those chronic conditions or diseases that contributed to at least 10% of the total costs and based on this rather arbitrary rule, we included the following in D: heart disease (H), diabetes (D), hypertension (HT), high cholesterol (HC), osteoarthritis (OA), asthma (A) and morbid obesity (O).

The flow diagram (14) for our simple model describes people acquiring new conditions and moving to other healthstates. (See Figure 1) N, D, and C stand for the population corresponding to the three distinct healthstates in the model. D includes people with both single and multiple co-morbid conditions. The arrows in the flow diagram represent the transitions of people from one healthstate to another and the number of people making the transitions is described using fractions.

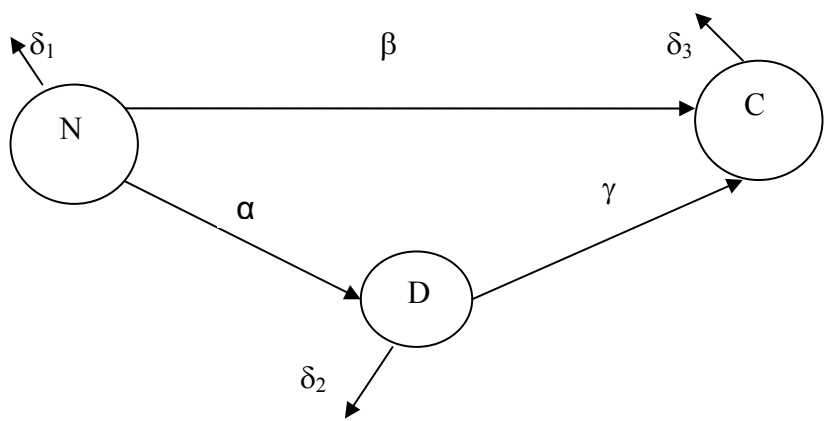

**Figure 1. Flow diagram for the three healthstate model.**
N, D and C are the three healthstates. α is the rate at which non-chronic patients acquire chronic disease; β is the rate at which non-chronic patients develop cancer; and γ is the rate at which chronic patients develop cancer. The three δ s describe the rate of drop-outs from the three healthstates.

In this three severity level model we have three distinct transitions: non-chronic people acquiring chronic conditions; non-chronic people acquiring cancer; and chronic patients acquiring cancer. Our calculations are for a period of one month and the corresponding *transition parameters* for the processes in the flow diagram are defined as follows:

- α   is the fraction of N that acquire one or more chronic conditions
- β   is the fraction of N that acquire cancer
- γ   is the fraction of D that acquire cancer

We analyze multiple years of claims data, and during this time there will be people leaving and entering the population segments for various reasons and at different times. These numbers used in the flow diagram are defined as follows:

- $\delta_1$  is the fraction of non-chronic patients that drop out
- $\delta_2$  is the fraction of chronic disease patients that drop out and
- $\delta_3$  is the fraction of cancer patients that drop out.

When we consider the more complex model with multiple chronic healthstates, flow diagrams tend to become confusing and, instead, we will represent them with the following transition table representation that describes all the transitions in an equivalent manner. (See Table 1).

| From healthstate    >> | N | D | C |
|---|---|---|---|
| To healthstate D | α | | |
| To healthstate C | β | γ | |
| Drop outs | $\delta_1$ | $\delta_2$ | $\delta_3$ |

**Table 1.**
The list of parameters for transition from healthstates N, D and C to the higher healthstates.
For example,  β is the fraction of N that transitions to C.

The differential equations corresponding to the flow diagram and Table 1, and describing changes in N, D and C during a time interval *dt* can then be written as follows:

$$\frac{dN}{dt} = -\alpha N - \beta N - \delta_1 N \quad (1)$$

$$\frac{dD}{dt} = \alpha N - \gamma D - \delta_2 D \quad (2)$$

$$\frac{dC}{dt} = \beta N + \gamma D - \delta_3 C \quad (3)$$

where N, D and C are all time dependent.

We can introduce a 3 x 1 state vector $\vec{I}(t)$ that describes the population at a time t as:

$$\vec{I}(t) = \begin{pmatrix} N(t) \\ D(t) \\ C(t) \end{pmatrix}$$

where $N(t)$, $D(t)$ and $C(t)$ are the three healthstates representing the number of non-chronic, chronic disease and cancer patients at a specific time "t".

The three equations of the dynamic system can then be simplified into a matrix equation:

$$\frac{d\vec{I}(t)}{dt} = L\,\vec{I}(t) \qquad (4)$$

where $L$ is a 3 x 3 matrix.

For our analysis it is more convenient to consider changes that take place in $\vec{I}(t)$ for a finite interval of time, say one month, rather than an infinitesimal interval dt, in order that the changes that take place in the components of $\vec{I}(t)$ are statistically significant and measurable. Once a finite interval of time is selected then the differential equation (4) takes the form of a difference equation.

The difference equation connecting a state vector at time "t" to one at "(t+1)" is then written as

$$\vec{I}(t+1) = M\,\vec{I}(t) \quad (5)$$

where $M = \mathcal{I} + L$, with $\mathcal{I}$ being the identity matrix. $M$ captures the dynamics of the process as the system evolves from time t to (t+1). In particular, the matrix elements of $M$ describe the rate of transition of patients from one healthstate to another and so $M$ is often referred to as a "transition matrix."

## 2. Matrix M and the phenomenology of population dynamics

Before attempting to solve eq. 5, it is useful to understand the structure and other details of the matrix $M$ and relate its matrix elements to processes that take place in a chronic population. These are calculable from claims data and they offer a phenomenological interpretation of chronic population dynamics that can be helpful as we try to find approximate solutions to eq. 5.

By definition, there is no cure for chronic conditions, and so the transitions in this model are always restricted to be unidirectional, from the least severe healthstate to the higher severity states. In other words, people belonging to $C$ do not ever transition back to $D$ or $N$, and people in $D$ do not get back to state $N$. This restriction is already contained in the differential equations and places a structural restriction on the matrix $M$.

If we keep in mind that the matrix element $M_{ab}$ refers to the transition from healthstate "b" to "a" then we see that the unidirectional transitions then implies that certain matrix elements ( 3 to 2, 3 to 1 and 2 to 1) are zero and $M$, therefore, takes a (lower) triangular form.

$$\begin{pmatrix} M_{11} & 0 & 0 \\ M_{21} & M_{22} & 0 \\ M_{31} & M_{32} & M_{33} \end{pmatrix}$$

From the flow diagram it is possible to deduce the various non-zero matrix elements of $M$ and they are:

$M_{11} = (1 - \alpha - \beta - \delta_1)$
$M_{21} = \alpha$
$M_{22} = (1 - \gamma - \delta_2)$
$M_{31} = \beta$
$M_{32} = \gamma$
$M_{33} = (1 - \delta_3)$       (6)

From healthcare claims data it is then possible to calculate all the elements of the matrix M just by counting the number of people whose diagnoses codes change during a selected period and transition to higher severity healthstates.

To solve eq. 5 a critical assumption that is made is that the transition parameters or the matrix elements of $M$ are *constant in time*. With this assumption it is possible to obtain the solution:

$$\vec{I}(t+n) = M^n \, \vec{I}(t) \quad (7)$$

Since all the non-zero matrix elements of $M$ are known, by repeated operations of $M$ it is possible to compute the changes that take place to the populations in healthstates *N, D,* and *C* in unit intervals of time.

As preventive care programs become more effective, there will be fewer transitions, and it is this inverse relationship that can be used as a metric for performance of the programs.

The variables $\delta_1$, $\delta_2$, $\delta_3$ - the net efflux of patients from each of the three population classes – may be influenced by external causes and their effects on predictions will vary from case to case. For Medicaid populations, enrollment qualifications are based on income, and that will change as people find or lose jobs. Clearly in such cases $\delta_1$, $\delta_2$ and $\delta_3$ are exogenous and can affect the model's predictions. For Medicare populations, on the other hand, the loss in membership is mostly due to mortality arising from the disease state, and so $\delta_1$, $\delta_2$ and $\delta_3$ are mostly attributed to endogenous reasons and pertain to the quality of care for the various disease states of the system. Commercial insurance enrollees have the option to switch insurers during open enrollment periods (usually twice a year) and so the net changes in enrollment during these periods can be unusually large and since these are exogenous, they cannot be accounted for in the model. In short, when these variable parameters are external to the model then the assumption that they remain constant in time may be violated and some explicit adjustments must be made to accommodate the variations in the value of the $\delta$s.

## 3. Model Calculations of Population Trends

In this section we will use eq. (7) to forecast the changes that take place in the three population segments over time and then compare predictions to actual observations obtained from claims data. Healthcare claims data include the diagnoses of patients and they can be used to assign and count the population in the segments, *N, D* and *C* at any specific time. We use a month as our unit of time step in the equation, and by counting the number of people acquiring a chronic condition and moving from *N* to *D* and moving from either *N* or *D* to C, we are able to measure the transition parameters in the matrix M. And once we know the vector $\vec{I}(t)$ and the matrix M, then from eq. (7) it is clear that by repeated application of the matrix M we can predict new values of *N, D* and *C*, one step at a time.

The approach then is to calculate the matrix elements using a subset of data and then predict the values of *N, D* and *C* for subsequent periods and compare these forecasts with the remaining actual claims data treated as observations. If the predictions match the observations then the validity of the model is established. Our objective was to compare a 12 month-forecast with observations.

The method consists of four steps. The first step, is to calculate the transition parameters of the matrix M. One month's healthcare claims data is all that is needed to derive the elements of the transition matrix M. But we used a three month average to get a better representative value of the parameters $\alpha$, $\beta$, $\gamma$, $\delta_1$, $\delta_2$ and $\delta_3$ . We used commercial insurance data and, as noted in an earlier section, enrollees tend to join and leave plans during the open enrollment period. But most insurers are loath to accept patients with pre-existing conditions, and so it is the non-chronic enrollees that usually take advantage of the open enrollment season. In other words it is the parameter $\delta_1$ that shows a large variation during the enrollment season. Our sample data confirmed the fact that the other parameters varied very slightly during the year, and the assumption that the matrix M is at worst, weakly time dependent seemed credible.

The second step is to create the population column vector $\vec{I}(t)$ obtained by calculating from the claims data, the three components *N*(t)*, *D*(t) and *C*(t), the total non-chronic, chronic and cancer patients at the start of the first month.

The third step is to apply M repeatedly, to calculate predictions of the populations for *N, D* and *C* for subsequent months and compare them to the actual data from claims for the same number of months.

In our analysis we used 39 months of de-identified commercial claims data (covering about 123,000 lives). We divided the data into four parts. The first was a 3 month period at the start that was used to calculate mean values of the model transition parameters at the outset. The remaining data were divided into 3 one-year "select" periods of *observations* and so we had three opportunities to check model forecasts against observations. For the second and third select periods we used the claims data from the previous period to recalculate twelve-month mean values of the transition parameters and a new transition matrix M.

Here are some of the sample graphs that describe how well the predicted calculations match the observations. We display the observed and predicted values of the number of people with chronic conditions during 11 months (See Figure 2). The error in prediction increases with time but even after eleven months the error in the predicted $D$ population is less than 3%.

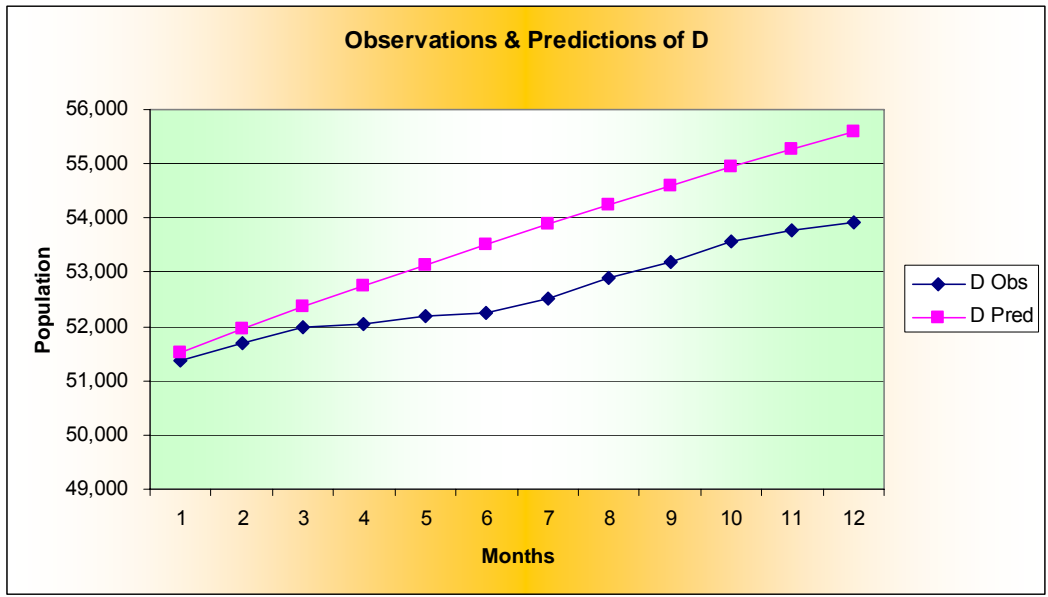

**Figure 2. Comparison of forecast of chronic population with observations.**
Predictions of the chronic population ($D$) obtained from the dynamic model are compared to the observed values.

We display the similar comparison of observations and predictions of cancer population during a 11 month period (See Figure 3). Here again the errors in the estimate increases with time. Even after eleven months, the error in the prediction of Cancer populations is less than 2% in this sample data set.

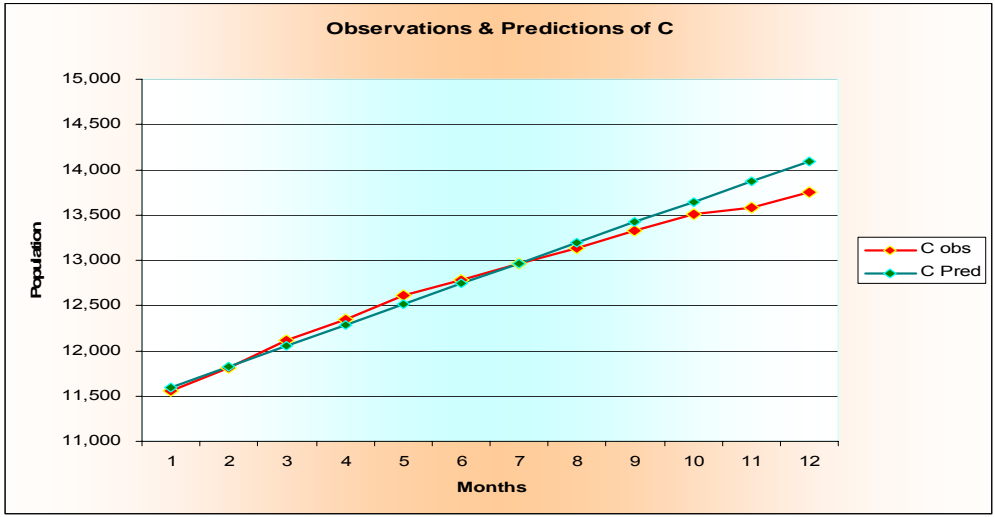

**Figure 3. Comparison of forecast of cancer population with observations.**
Predictions of the Cancer population obtained from the dynamic model are compared to the observed values during a 11 month period.

We have chosen to display the results from one selected period since the two other select periods also yielded similar results with low error values. We have displayed the error estimates in *D* and *C* for the three select periods of data that we used for validation of the approach (See Figure 4).

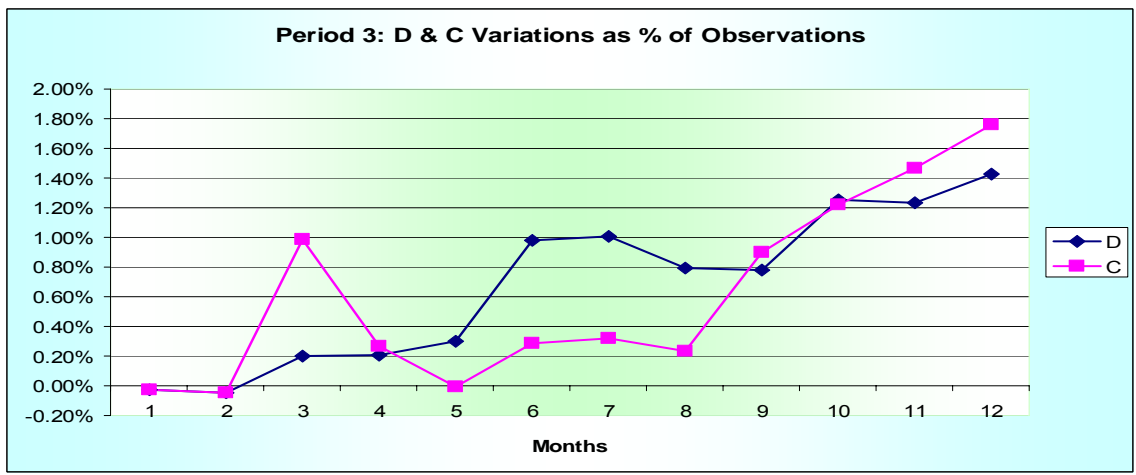

**Figure 4. Errors in the forecasts as a percentage of the observations.**
We display the percentage variation of predictions from observations for both chronic disease and cancer populations for one of the three cases considered. In all three cases the estimates in the chronic population segment had a less than 4% error while the error in estimating cancer patients was less than 3%

The graph in Figure 4 displays the variations in the % error in estimation of chronic and cancer patient populations for the 11 month period as a fraction of the observations. The quantities plotted are

$$\Delta D\% = \frac{D_{pred} - D_{obs}}{D_{obs}} \text{ and } \Delta C\% = \frac{C_{pred} - C_{obs}}{C_{obs}}$$

for the three sets of data, where the subscript "pred" stands for "predicted" and the subscript "obs" for "observed". The maximum error in $\Delta D\%$ *and* $\Delta C\%$ is less than 4% and 3% respectively, confirming that the predicted values are in good agreement with observations. This provides a general validation of the three healthstate population model.

We repeated the analysis for the population segmented into age groups and the results obtained are similar, with forecasts matching observations with less than 5% error.

If there are wellness programs that target the non-chronic population and they are effective, then the trend in acquisition of chronic diseases will decrease and the corresponding transition parameters will decrease. In other words, the transition parameters can be used as a metric to compare effectiveness of preventive care programs.

## 4. Multiple Healthstate Model

We have assumed a simple three healthstate model and there is a price we pay for simplicity. The healthstate D consists of people with different chronic diseases, with varying medical needs and costs based on the severity of their diseases. Even though the health status of the chronic patients in the population is constantly changing as people acquire more chronic diseases, the model assigns them to a single healthstate. In the same vein, our choice of a single healthstate C for cancer, limits our ability to distinguish the relationship of chronic diseases to various types of cancers in the population.

The limitations of our simple model can be addressed by choosing a different basis to describe our population by expanding the number of healthstates appropriately. The more complex version of the model we propose segments our sample population into six healthstates that correspond to those people with no chronic conditions; 1 chronic condition; 2 chronic conditions; 3 chronic conditions; more than 3 chronic conditions; and lastly, those with cancer. The number of people with more than 3 chronic conditions was relatively small and so we assigned them all to single healthstate and described it as those with more than 3 chronic conditions.

The chronic conditions selected were the same seven identified in the three healthstate model and based on cost. Applying the same rule to other sample populations may yield a different set of chronic conditions. An alternate and equally valid rule would have been to consider only those chronic conditions that affected at least a specific percentage, say 10%, of the population.

There will be many more transitions in this model, and the forecasting ability of the model will be put to an even more severe test as it attempts to predict the changes that take place in the multiple healthstates during the course of twelve months.

As in the three-healthstate model, transitions occur only in one direction, from those with lower numbers of chronic conditions to higher ones, with cancer occupying the healthstate with the highest severity of disease. The flow diagram can be complex and confusing and so we will use instead the transition table formalism (as in Table 1) to depict the various transitions. Adapted to the multiple healthstates the transitions can be identified directly (See Table 2).

| From healthstate    >> | N | 1d | 2d | 3d | >3d | C |
|---|---|---|---|---|---|---|
| To healthstate 1d | a | | | | | |
| To healthstate 2d | b | f | | | | |
| To healthstate 3d | c | g | j | | | |
| To healthstate >3d | d | h | k | m | | |
| To healthstate C | e | i | l | n | p | |
| Drop outs | $\delta_0$ | $\delta_1$ | $\delta_2$ | $\delta_3$ | $\delta_4$ | $\delta_c$ |

**Table 2.**
The list of transition parameters from one healthstate to a higher healthstate. For example, g is the fraction of people in healthstate 1d that transition to 3d in the selected period.

Using a scheme analogous to the three healthstate model we define the transition parameters in Table 2 for a selected period as:

a is the fraction of N that transitions to 1d
b the fraction of N that transitions to 2d
c the fraction of N that transitions to 3d
d the fraction of N that transitions to >3d
e the fraction of N that transitions to C

f the fraction of 1d that transitions to 2d
g the fraction of 1d that transitions to 3d
h the fraction of 1d that transitions to >3d
i  the fraction of 1d that transitions to C

j the fraction of 2d that transitions to 3d
k the fraction of 2d that transitions to >3d
l the fraction of 2d that transitions to C

m the fraction of 3d that transitions to >3d
n the fraction of 3d that transitions to C

p the fraction of >3d that transitions to C

and the fractional drop outs from healthstates N, 1d, 2d, 3d, >3d and C are $\delta_0$, $\delta_1$, $\delta_2$, $\delta_3$, $\delta_4$ and $\delta_c$ respectively. The physical interpretation of the matrix elements and the phenomenological explanations continue to be valid in this version as well.

As in the three healthstate model, we considered three sets of data and in all three sets the predicted populations in the various healthstates are quite close to observations. We display a representative sample of the results from the third set of data (See Figure 5). The variations in prediction of populations in the 1d, 2d and C healthstates as a percent of the values of the corresponding observational data are displayed here. Errors continue to be small in this model as well - less than 2% for the entire period.

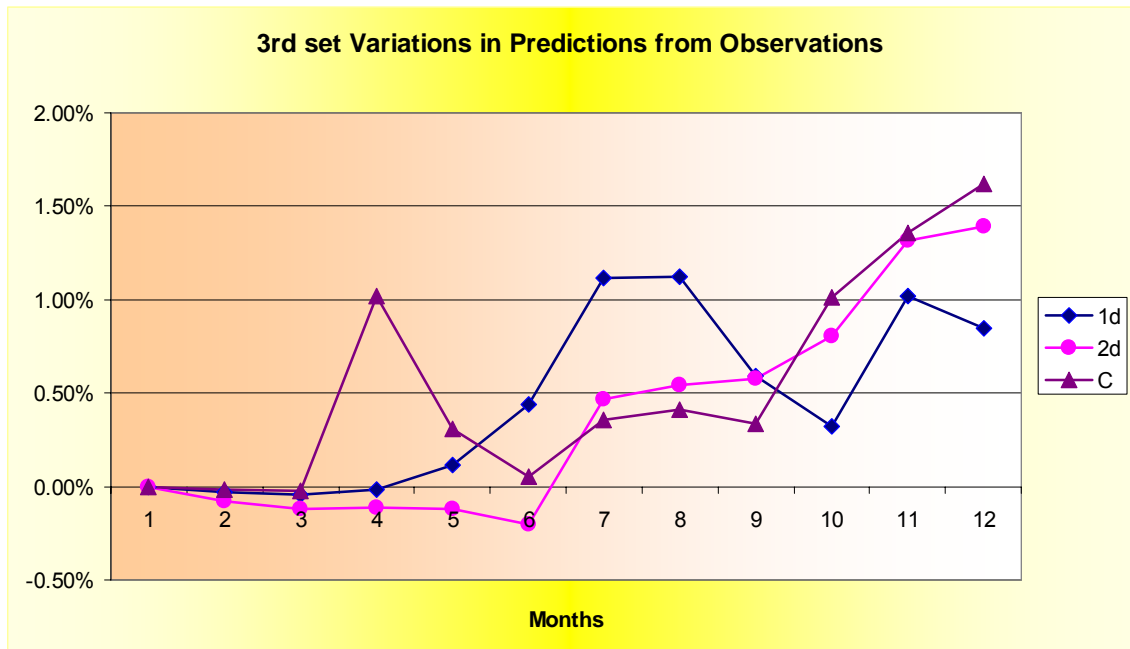

**Figure 5. Errors in forecasts of various chronic healthstates and cancer.**
Percentage variations of calculations of 1d, 2d and C populations from observations for a twelve month period are displayed in the figure. To avoid clutter we have excluded the other healthstates in this figure.

The maximum errors in prediction of the population of various healthstates compared to observations can then be calculated (See Table 3). The definitions of the various quantities are similar to the earlier three healthstate model. For example

$$\Delta 1d = \frac{1d_{pred} - 1d_{obs}}{1d_{obs}}\% \text{ and } \Delta C = \frac{C_{pred} - C_{obs}}{C_{obs}}\%$$

|  | Maximum Error in Predictions | | |
|---|---|---|---|
| Healthstate | Set 1 | Set 2 | Set 3 |
| $\Delta 1d$ | -1.11% | -2.39% | 0.85% |
| $\Delta 2d$ | 2.84% | 0.30% | 1.39% |
| $\Delta 3d$ | -2.07% | 2.94% | 0.98% |
| $\Delta{>}3d$ | 3.43% | 6.08% | 1.70% |
| $\Delta C$ | 1.35% | 3.12% | 1.62% |

**Table 3.**
The maximum deviations in calculated or predicted population of healthstates from the observed values, as a percentage of the observed values, are displayed in the table for the three sets of observational data.

All the errors are less than 3.12% except for the healthstate >3d, where the error is 3.43% and 6.08% for two of the data sets. Error here is larger presumably because the healthstate represented as >3d includes patients with 4, 5, 6 or 7 disease conditions and so different numbers of multiple types of transitions are being treated as if it were a single transition category.

Once again, as in the three healthstate model, the number of transitions from a healthstate is inversely related to the effectiveness of preventive care programs. Transitions from the non-chronic state measure the performance of what are called "wellness programs", to keep a person from acquiring a chronic condition. Patients who have a single chronic disease are usually managed by primary care physicians and so the transitions away from 1d are a measure of the efficacy of their management. Those with several chronic conditions may consult multiple specialists and the transitions from higher healthstates will reflect the effectiveness of the coordination of care.

In the multi-healthstate model, costs for chronic conditions appear in much greater detail. There are two costs that are important: the per capita cost for each healthstate and the additional costs incurred due to transitions from each healthstate to higher healthstates. These costs, together with the population density in each healthstate and the number of transitions from one healthstate to another provide a complete description of the population and the associated costs. We can also create geographic maps that we call "Care Maps", showing both disease burdens and costs and their trends from the results obtained through our analysis. Discussions of Care Maps" are beyond the scope of this paper.

## 5. Population Analyzed in Terms of a Single Disease

While the multiple healthstate model looks promising it is not clear if epidemiologists will be able to adapt their analysis to include co-morbid conditions easily. Also, there is a considerable amount of research based on single disease analysis and not being able to connect with them would leave our population model isolated and at a disadvantage. There are surveys of disease trends and the maps created from them invariably deal with single diseases. Many investigations of cancer risk focus most of the time on a single chronic condition like diabetes or heart disease and some of them have discovered a correlation between a chronic disease and increased cancer risk. In this section we will show that we can use our model for single disease analysis also, and prove this by duplicating the results obtained from epidemiological investigations.

For example, in a paper written by the Emerging Risk Factor Collaboration (16), the authors concluded that diabetes patients are at 25% higher risk of mortality from cancer. Insurance actuaries track costs carefully and they know that cancer costs are 10 or more times as much as those for chronic diseases, and so their computed premiums must reflect this increased risk and be unusually high for diabetes patients. But that is not the case. The premiums charged by insurers for diabetes patients are about the same as those for patients with hypertension or high cholesterol. It's puzzling that epidemiologists detect an increased clinical risk for diabetes patients but actuaries find that the financial risk posed by diabetes is no worse than those of other chronic risk factors. Our model offers an explanation to this puzzle.

To perform single disease analysis we need to change our basis healthstates. We will segment chronic patients into those with and without diabetes, and so in this new formulation we will have four healthstates to describe patients: non-chronic (N); those with diabetes and other co-morbid conditions (D+); those with chronic and co-morbid conditions but not diabetes (D~); and those patients with cancer (C).

The flow diagram can be drawn from the following transition table (See Table 4).

| From healthstate    >> | N | D~ | D+ | C |
|---|---|---|---|---|
| To healthstate D~ | c | | | |
| To healthstate D+ | a | f | | |
| To healthstate C | b | e | d | |
| Drop outs | $\delta_0$ | $\delta_1$ | $\delta_2$ | $\delta_c$ |
| | | | | |

**Table 4.**
Transition table for the single disease healthstate model. The four healthstates are N for non-chronic, D~ for those chronic states without any of the specific single disease we are studying and D+ stands for the healthstate that includes the specific single chronic disease and other co-morbid states and C stands for Cancer.

As before, we will use part of the data to calculate the new transition parameters and retain the rest of the data as "observations". Once again we find that the forecast of cancer rates is good, with less than 5% error.

What is even more interesting is that, unlike epidemiological analysis, our population based formalism makes it easy to study various other chronic conditions rapidly and without much extra effort. In particular, we can perform a similar analysis for the various other chronic states in our sample data, and thus compare cancer incidence rates for different chronic risk populations.

A quantity of interest for this analysis is called "conditional probability" P (C|D) - the probability of a person with diabetes acquiring cancer - and that can be calculated from the rate of incidence of cancer from healthstate D+, and this can be derived from claims data. We can similarly calculate the conditional probability of a non-chronic patient acquiring cancer, P(C|N).

For comparing risks of various chronic diseases it is useful to consider the quantity $\dfrac{P(C \mid D)}{P(C \mid N)}$,

which we call the normalized conditional probability of diabetes patients acquiring cancer.

As epidemiologists do, we study distinct age groups, in this case people in age ranges 50-59 and 60-69 separately. We calculate the normalized conditional probability for each of the chronic conditions and display them in a table (See Table 5). We find, as did others, that those with diabetes are indeed at a higher risk for cancer, for both age groups. Additionally, we discovered the unexpected result that those with any of the other chronic conditions also had a similar higher risk for cancer. In other words, no matter which chronic condition was present, there was an increase in cancer incidence, and the risk was about 1.2 and 1.5 times that in the non-chronic case.

|  | P[C\|H] / P[C\|N] | P[C\|D] / P[C\|N] | P[C\|HC] / P[C\|N] | P[C\|HT] / P[C\|N] | P[C\|A] / P[C\|N] | P[C\|OA] / P[C\|N] |
|---|---|---|---|---|---|---|
| 0 - 59 | 1.505±0.002 | 1.223±0.006 | 1.283±0.011 | 1.228±0.008 | 1.444±0.011 | 1.292±0.007 |
| 0 - 69 | 1.383±0.035 | 1.178±0.036 | 1.219±0.041 | 1.212±0.041 | 1.426±0.010 | 1.238±0.036 |

**Table 5.**
We have calculated the various conditional probabilities normalized to the non-chronic case for the age ranges 50-59 and 60-69.

It is important to emphasize that this result is obtained from a direct analysis of the sample data. Of course, our model will also yield the same result since we are able to predict cancer risk with a good accuracy for various single diseases. For the statistical analysis we also segmented the population into various healthstates as prescribed by our dynamical model. The results of Table 5 are displayed as a graph. (See Figure 6.)

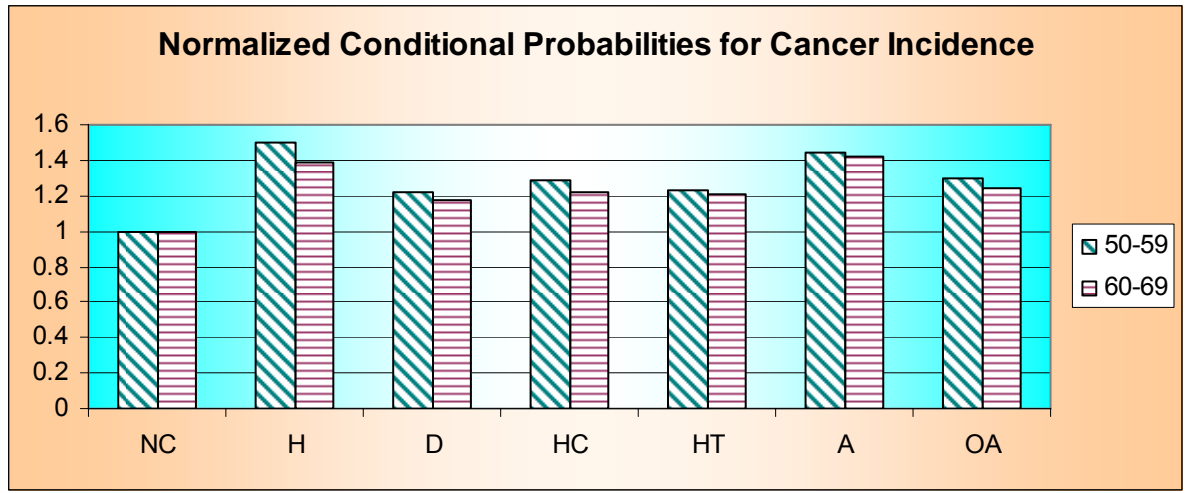

**Figure 6. Cancer risk for various single disease types.**
The results displayed in Fig. 6 are obtained from directly analyzing the claims data but using the healthstate segmentations in our model to focus on a single disease at a time. Cancer risk seems to increase with the presence of any chronic disease, and in this sample data it seems to be 1.2 to 1.5 times higher when compared to those with no chronic diseases.

Our result agrees with the conclusion of the ERFC study (16) that diabetes increases cancer risk, but our data also seems to vindicate the actuarial experts at insurance companies who equate the financial risk of diabetes to those of other chronic conditions. From a public health planning point of view, this is an important distinction. If, for example, based on the ERFC results it is decided to focus solely on diabetes prevention to lower cancer risk, then our results show that this may be counter-productive, since other chronic risk factors that are ignored may increase the cancer incidence rates in the population.

Parenthetically we must note that an epidemiological analysis for each of the separate chronic conditions would require a considerable investment in time and resources and that is probably why such studies have not been conducted previously. Our population model based approach is rapid and once a single disease analysis is completed then the results for other diseases are obtained from trivial tweaks of the queries.

Of course, the general validity of this result that any chronic disease increases the risk of cancer must be verified independently, with other sets of data. But this result also raises the converse issue of great practical importance: will effective prevention of chronic disease automatically reduce the cancer population? We can answer this question using our three healthstate model.

The N to C and D to C are direct transitions governed by the parameters $\beta$ and $\gamma$ and programs like smoking cessation and lowering obesity rates are designed to effectively reduce their magnitude, and then clearly the incidence rates for cancer will also drop. $\gamma$ for example is of the order of $10^{-3}$ for our sample data and so the impact on incidence rate is approximately 0.1% of the population in D(t).

But will chronic preventive care programs like, say diabetes or hypertension prevention, lower cancer incidence rates? In the three healthstate case we are trying to understand if lowering of the value of the parameter $\alpha$ from typical wellness programs will indirectly reduce cancer incidence rates. Actually in our model, the effect of $\alpha$ on C(t) is a second order effect, influencing it through the changes in D(t). In our sample data $\alpha$ contributes a very small change in D(t) (less than 1%) and the resultant impact on C(t), using the result in the previous paragraph, is further reduced by a factor of the order of $10^{-3}$, making it a very small number.

The conclusion applies only to the sample data but our result seems to suggest that simply improving the effectiveness of chronic preventive care programs will only have a tiny effect on the incidence of cancer in populations. The parameters $\beta$ and $\gamma$ are the ones that directly affect C(t) and lowering them can affect cancer incidence rates and they are the goals of typical cancer prevention programs.

## 6. Savings from preventive care programs: myth or real?

It's an article of faith amongst clinicians that preventive care programs will save money. Companies that provide "wellness" programs also assert, with a touch of self interest, that investments in preventive care will generate large returns for employers. The federal government and insurance companies have gone even farther and instituted projects where providers get part of their payments as a share of the "savings" arising from successful preventive care efforts.

Yet there are critics who question whether there ever will be savings, and their compelling logic is based on simple arithmetic. Preventive care programs will require more office visits and a greater frequency of lab tests and this will increase total costs of care. Further, since we are unable to identify the few people that can benefit most, we are forced to extend preventive care to everyone in the population. As a result, during any period, the additional cost of the programs will be much higher than the reductions obtained by keeping a few people from becoming chronically ill. This then is the basis of the observation (17, 18) that the net effect over multiple periods will only further raise healthcare costs and there will be no savings.

Cost justification is possible even when there are no savings, especially if care programs prevented catastrophic events in a few people. But to prove that an event didn't occur because of the preventive care program is not easy. After all, there was never a certainty of occurrence in the first place, and so the non-occurrence of an event could have been by chance. In short, justification of investments in expensive care programs becomes difficult when there is neither savings nor proof of benefits.

Our model helps address both these concerns. First we expose a subtle arithmetical flaw in the reasoning that suggests there cannot be any savings. Instead of focusing on catastrophic events, we prove that preventive care programs have the potential to improve the health of populations and they are manifested as reductions in transitions to higher healthstates. Our calculations show that these savings may take too long to materialize, and so they are best justified as a precautionary investment to mitigate future financial risk.

To digress for a moment, it is necessary to establish baseline costs of a population so as to calculate savings. One useful measure is a "per capita cost" but that is problematic when there are people with varying severity of diseases and treatment costs. Just a few people with catastrophic disease conditions can then skew the mean values considerably and render them virtually useless for comparisons. We step around this objection in our model, by segmenting populations into healthstates that have a more uniform severity of disease, and hence assign people with similar medical needs to the same group. We are then able from claims data, to calculate per capita cost for each healthstate as well as the increase in costs as people transition from one healthstate to another. Of course, we are only referring to "direct" costs. There are additional societal and individual costs arising from complications of chronic diseases that are not included in this analysis and are beyond the scope of this paper, such as poor quality of life, increased sick days, and decreased worker productivity.

If a patient transitions to a higher healthstate then, because chronic conditions tend to persist, the person will continue to incur more costs in the new healthstate, and the baseline costs of the population for the future periods will have risen. The critical factor in our analysis is the converse: if a preventive care program reduces the number of transitions in a period, not only will there be a cost reduction for the first period, the benefits of the reduction continue and the baseline costs in all future periods will also have reduced by this amount.

We can derive a mathematical expression for break-even, when increased costs from preventive care programs equal the cost reductions arising from lowering of transitions as follows. If we assume $\Delta S_j$ is the increase in costs in a healthstate for the $j^{th}$ period due to the preventive care program, and $\Delta T_j$ is the corresponding reduction in costs arising from the lowering of transitions in the same period, then we can calculate the increased costs and the reduction in baseline costs in each period as follows:
In period 1, the increase in costs is $\Delta S_1$ while the baseline cost reduction is felt in "n" subsequent periods and is, therefore, $n\Delta T_1$.
In period 2, the increase in costs is $\Delta S_2$ while the baseline cost reduction of $\Delta T_2$ is effective in "n-1" subsequent periods and so on.
If we now add the total increase in costs for the n periods and equate it to the total reduction baseline costs then we will be able to calculate the break even value of n:

$$\sum_{j=1}^{n} \Delta S_j = \sum_{j=1}^{n} (n-j+1)\Delta T_j \quad (8)$$

Notice that the left hand side is a simple sum while the right is actually the sum of an arithmetic series. This is the mathematical explanation of why, even though in each period the preventive care programs cost more than the reductions in cost from lowering of transitions, the reductions continue to take effect in subsequent periods and eventually exceed the additional costs attributed to preventive care programs – and there will be savings. Does it then make sense to offer a share of the savings to providers, as insurers and the federal government propose, as an incentive to promote preventive care programs?

To answer this question we need to get a better understanding of the savings, and so let us make the simplifying assumption that, for each period, the additional investments and the reduction in costs are constant and are respectively ΔS and ΔT. Then we can sum the arithmetic progression on the right to get

$$\Delta S = \frac{(n+1)}{2} \Delta T \quad \text{or equivalently,}$$

$$n = 2\left(\frac{\Delta S}{\Delta T}\right) - 1 \quad \text{eq. ( 9)}$$

Using our data and some representative values for cost and savings, we will be able to calculate values for $n$, the break-even period. We will assume that $20 per member per month is the additional cost incurred due to the preventive care program. Employers at this time pay between $500 and $1000 per employee for wellness programs annually and so our estimate is a reasonable and low one. Let's assume it produces a 20% reduction in the number of transitions for the various age ranges and that may be a little difficult to reach. In making these assumptions we will obtain a low estimate of $n$ and the actual value of $n$ in reality will be much higher.

It is then possible to calculate the value of $n$, the break-even number of months, using equation ( 9) and they are listed in the table (See Table 6) for various age ranges.

| Age range | $n$ (in months) |
|-----------|-----------------|
| 30-39 | 124 ± 7 |
| 40-49 | 72 ± 5 |
| 50-59 | 48 ± 4 |
| 60-69 | 39 ± 3 |

**Table 6.**
Assuming the increase in cost per person per month is $20 for preventive care, and that results in a 20% reduction in the number of transitions, the number of months $n$ it takes for them to be equal is listed in the table for various age ranges.

The assumptions we have made are rather simple but it does give us a rough estimate of when savings will begin to materialize. For example, in the age range 40-49 where there is usually the largest number of people vulnerable to chronic illness, it takes 72 months for savings to set in. Of course, the actual values of $\Delta S_j$ and $\Delta T_j$ must be used to find the exact break-even period, but it is obvious that several years may pass before such programs yield a positive return.

In the long term there will be savings but providers have immediate expenses to meet and cannot afford to wait that long to get their share. Employers and insurers are not going to rush to invest when yields are so far in the future. In other words, these savings do not provide a sufficient and timely incentive for providers and neither do they make a compelling case for investments in preventive care programs.

But there is another way to justify investment in preventive care programs and that is to view it as a precaution against increasing future financial risk. Seat belts and bicycle helmets are examples of investments made to mitigate risk of bodily harm. So also, factories invest in sprinkler systems as a protection against fire hazards. Such investment decisions are decided using actuarial models, without any consideration of savings. Healthcare can follow this path and adopt a new paradigm for cost justification based on risk-mitigation.

Once viewed as a risk, actuaries can decide how much of an investment in prevention is warranted to hedge against this risk (19). Based upon the program's goals, they can now calculate incentives for providers to offer better quality of care as evidenced by the lowering of both the number of transitions and the break-even period. This will serve the dual purpose of reassuring providers that their efforts will be rewarded when they meet realistic goals and at the same time convince employers and insurers that preventive care investments are part of normal business costs and an essential component of prudent risk management.

While the focus of the above analysis has been on wellness programs that prevent non-chronic patients from acquiring chronic diseases and the associated transitions from non-chronic health states, parallel arguments apply for other healthstate transitions as well. Limiting 1d state transitions can then be used to provide incentives for the primary care physician while 2d and higher state transitions can provide a similar measure for specialists involved in the coordination of care.

Summarizing, our model shows that there are two components to costs. The first we identified as the per capita costs for a healthstate. These costs are attributed to office visits, lab tests, drug costs and, as a rule, preventive care programs will increase these costs. Because we are dealing with chronic diseases with no cure, these costs of treatment contributing to the first component will continue unabated. The second component is the additional and future costs arising from healthstate transitions, and they will gradually increase baseline costs. Effective prevention and clinical care programs will lower the numbers of transitions and will help reduce the future rise in baseline costs. Our model can split costs into these two components and calculate them separately and the consequent mitigation of risk can be the basis of a new paradigm to measure provider performance and reward providers who meet performance goals.

## *Discussion*

We have presented a model that provides a rapid, inexpensive way to measure and forecast the health and healthcare costs of a population, and that can be used as a metric for improvement in both areas. Public health officials, insurers and government, will be able to use these forecasts to plan and allocate their clinical and financial resources in an optimal manner to better serve the communities.

Big data (20, 21) refers not only to the volume of data but also to the variety of data, both structured and unstructured, and also the speed with which they accumulate and change. It offers the potential for gaining insights through its advantages in handling the gray data prevalent today and helping make decisions faster. In the case of population health we have structured data from claims that contain demographic details, diagnoses, dates, cost and pharmacy information, and with increasing acceptance of electronic medical records, medical and laboratory data. But population health also includes results of surveys (22, 23, 24) and other sources of unstructured data about physical and socio-economic factors that affect the population, such as access to care, availability of nutritious food and open spaces that encourage physical activity.

Our contribution to big data of population health is in devising a metric and a unified analytics platform using the constructs of "healthstates" and "transitions" to quickly measure the health of populations. We are able to detect and measure improvements in health of populations whether it be due to new clinical interventions or changes that affect social determinants of communities. Mapping temporal and spatial variations in health of populations is very useful for multiple stakeholders. Eliminating ineffective programs and replicating successful ones are critically important for prudent management of financial and clinical resources of a community.

### *Acknowledgments*

One of the authors (JGK) would like to gratefully acknowledge Prof. Robert May's encouragement to carry our investigation beyond phenomenology and statistical correlations, and actually solve the equations of the dynamical system.  We are thankful to John Mendelsohn, at that time CEO of M. D. Anderson Cancer Center, who invited the author to meet with their senior staff, Professors Xifeng Wu, Ernest Hawk, Lewis Foxhall, Donald Berry and others, and the discussions that followed ignited the author's interest in seeking a mathematical explanation for the observed correlation of chronic diseases with cancer risk.   Both authors would like to thank the following for discussions at various times during the preparation of the paper: Professors Marianne Berwick, Jeffrey Griffith, Vasudev Kenkre, all of the University of New Mexico and Prof. J. D. Murray of the University of Washington.  At various times one of us (JGK) has had multiple discussions on the clinical and public health aspects of cancer and chronic populations, and would like to thank Aroop Manglick M.D., Reza Mehran M.D., Joseph Prendergast M.D., and Eduardo Sanchez M.D  We are also grateful for the discussions with Professors Edward Bedrick of the University of New Mexico, Jake Thomas of Yale University and Loren Cobb of the University of Colorado in Denver on various aspects of statistics as well as multiple linear and logistic regression relations.